Title:

# Angle-Resolved Thermal Emission Spectroscopy Characterization of Non-Hermitian Meta-Crystals

Fan Zhong[1,2,†], Kun Ding[3,4,†], Ye Zhang[2,†], Shining Zhu[2], C. T. Chan[3] and Hui Liu[2,*]

**Affiliation:**

[1] School of Physics, Southeast University, Nanjing 211189, China

[2] National Laboratory of Solid State Microstructures & School of Physics, Collaborative Innovation Center of Advanced Microstructures, Nanjing University, Nanjing, Jiangsu 210093, China

[3] Department of Physics, Hong Kong University of Science and Technology and William Mong Institute of Nano Science and Technology, Clear Water Bay, Kowloon, Hong Kong, China

[4] The Blackett Laboratory, Department of Physics, Imperial College London, London SW7 2AZ, United Kingdom

[†] These authors contributed equally.

[*] Corresponding author: liuhui@nju.edu.cn

**Abstract:**

We establish the angle-resolved thermal emission spectroscopy (ARTES) as a new platform to characterize the intrinsic eigenmode properties of non-Hermitian systems. This method can directly map the dispersion of meta-crystals within the light cone with a high angular resolution. To illustrate its usefulness, we demonstrate the existence of bound states in the continuum (BICs) and non-Hermitian Fermi arcs in a planar corrugated meta-crystal by measuring its angle-resolved thermal emission spectra. We show that change in the thickness of the meta-crystal can induce a band inversion between a BIC and a radiative state, and a pair of exceptional points emerge when the band inversion occurs. With this approach, the band mapping of non-Hermitian photonic systems can become a relatively striaghtforward task.

## I. Introduction

The angle-resolved photoemission spectroscopy (ARPES) has proved to be an important tool for measuring the energy and momentum of electrons in solids and has provided indispensable insights into the physics of many important materials, from superconductors to topological materials[1]. ARPES requires an external light source to excite and eject the electrons. In order to obtain a good resolution of energy and momentum, sophisticated and complex light sources are necessary. In physical systems like cold atoms and optical systems, band dispersion is typically determined through various optical excitation methods, including photoexcitation emission[2,3], reflection spectrum[4], and near-field microwave probe[5]. All these measurements also require external excitations, in which the external light source excites the eigenmodes inside the material, and the energy and momentum of the scattered photons are recorded to determine the band dispersion. Here, we demonstrate an angle-resolved thermal emission spectroscopy (ARTES) method which can perform band mapping without an external source. This method is simple, fast, and has excellent angle/frequency resolution, so that the measured "width" of the band is intrinsic and contains information about the coupling coefficient of the state with other degrees of freedom, such as the coupling to free space photons. As such, ARTES provides a natural platform to explore non-Hermitian physics.

The physics of non-Hermitian systems in photonics, as the generalization of parity-time (PT) symmetry[6], has attracted much attention recently. The realization is usually done by inducing gain and/or loss into photonic systems[4,7-18], leading to a wealth of potential applications based on exceptional points (EPs) [11,19-27]. Most finite-sized photonic structures are naturally non-Hermitian as they have radiative losses, but symmetry-protected radiationless states that can appear in many optical systems[28-37]. In addition, the recent development of topological concepts introduces extra degrees of freedom to non-Hermitian physics, and the interplay among geometry, symmetry, and topology is frequently associated with the emergence of EPs[38-47]. In particular, an EP can be regarded as a fractional topological charge, and hence the trajectory of eigenmodes joining the EPs of opposite chirality are sometimes called Fermi arcs[42]. EPs can also form connected structures, such as exceptional rings[43] and surfaces[44], in photonic structures. This shows that non-Hermiticity can lead to numerous interesting phenomena, and therefore it is highly desirable to investigate them experimentally.

Our ARTES setup can achieve an excellent angular resolution in the infrared region while maintaining simplicity in measurement. Compared to ARPES, a light source is unnecessary in constructing ARTES since the eigenmodes of the sample radiate naturally under thermal excitations. To demonstrate the ability of ARTES, a planar meta-crystal whose dispersions can be flexibly tuned by changing its structural parameters is designed and fabricated. By utilizing the symmetry of the structure and the radiation loss, novel non-Hermitian dispersion features such as BICs, bulk Fermi arcs, and exceptional lines (ELs) are investigated through ARTES in a synthetic parameter space for the first time. Both real and imaginary parts of eigenfrequencies can be obtained by the measurement.

**II. Angle-Resolved Thermal Emission Spectroscopy and Structure Design**

The experimental setup of the ARTES measurement is shown in Fig. 1(a). Here, our sample is placed in a heater (Linkam FTIR600) to keep the temperature at 100°C with a temperature stability within 0.1°C. The sample embedded within the heater is placed on a rotational stage with the minimum step of rotation being 0.2 degrees. We rotate the the sample around the z-axis and collect the emission from the sample through the heater window. The distance between the sample and the detector is about one meter, which is large compared to the wavelength, ensuring excellent angular resolutions. The thermal emission signal is analyzed using Fourier Transform Infrared Spectroscopy (FTIR, BRUKER VERTEX 70) at varied angles where a polarizer is used to assure only TE wave ($E_z$) is recorded by the FTIR. The working frequency was chosen around 30 THz in the atmospheric window, and the resolution of frequency is 0.03 THz ($1\text{cm}^{-1}$), which is high enough for the measurement. The band mapping can be realized without an external probe source. The simplicity and stablity of the ARTES will make it useful for characterizing applicable devices. The high resolutions, both in angle and frequency also ensure that the measured width of the band is instrinic and the real and imaginary parts of dispersion can be meaningfully extracted to provide insight into the non-Hermitian physics.

The design of meta-crystals with flexible tunning parameters is first discussed. The prototypes of the designed samples are shown in Fig. 1. The side-view SEM picture of one unit cell is shown in Fig. 1(b), and the corresponding schematics with structural parameters are given in Fig. 1(c). The thickness of the underlying Ge layer ($h_a$), the size of Au grating blocks (height $h_d$ and width $d_b$), and the gap width ($d_a$) between two adjacent Au blocks are fixed after photolithography. However, the thickness of the top Ge layer ($h_b$) can be continuously changed by depositing additional Ge on top of the Au grating, and hence $h_b$ can work as a tunable parameter. Figure 1(d) shows the top view of a small portion of the whole sample. The thermal emission of this meta-crystal is measured with ARTES. In this experiment, the ARTES measurement is carried out, and the $h_b$ is changed alternatively, thereby obtaining the thermal emission spectra with varied values of $h_b$.

In experiment, the structure is fabricated on a silicon wafer (2 × 1.6 cm) according to the design in Fig. 1(c). A 200 nm gold layer and a Ge layer with thickness $h_a$ are first deposited through electron beam evaporation (EBE). Afterwards, the gold grating is fabricated through photolithography with a width $d_b$, thickness $h_d$, and period $\Lambda = d_a + d_b$. The Ge layer is again deposited on the grating with the designed thickness $h_b$ through EBE.

### III. Numerical Simulations for BICs and EPs

We focus on the guided mode in the Ge layer of the designed structure in Fig. 1(b). Both the real and imaginary parts ($\text{Re}(n)$ and $\text{Im}(n)$) of the effective mode index of each multi-layer structure can be controlled by the geometric parameters of the cladding Au grating and Ge layers. The Au blocks significantly enhance the contrast of $\text{Re}(n)$, and the tunable parameter ($h_b$) can control the $\text{Im}(n)$ difference between these two modes. To guide the design of BICs and EPs, an effective one-dimensional meta-crystal model is established to describe the system, in which each multi-layer region is treated as an effective homogeneous layer with its effective mode index (see Section I of Supplementary Material). This model shows that the tuning parameters can induce the band inversion of BICs with a radiative state, leading to a pair of EPs connected by a non-Hermitian Fermi arc.

To reassure that the model provides a correct prediction before performing the experiment, the dispersion of TE polarized guided modes of the meta-crystal are investigated with different $h_b$ using full wave simulations COMSOL . The calculated dispersions of $\text{Re}(\tilde{f})$ for three $h_b$ are given in Figs. 2(a)–2(c), and the corresponding dispersions of $\text{Im}(\tilde{f})$ are given in Figs. 2(d)–2(f). A gap between two photonic bands is seen at $h_b = 0.78$ μm. The gap becomes smaller as $h_b$ increases. The gap closes at $h_b = 0.877$ μm and a pair of EPs appear. Further increase of $h_b$ opens the gap again, as shown in Fig. 2(c). In order to give a holistic view of the evolution of dispersions, we plot in Fig. 2(g) the real parts of eigenfrequencies $\text{Re}(\tilde{f})$ in a two-dimensional synthetic parameter space $(k, h_b)$. The color of this surface represents the inverse of imaginary parts of eigenfrequencies $\text{Im}(\tilde{f})$. A BIC line at *k*=0 can be clearly seen, which is mandated by symmetry. At $h_b = 0.78$ μm, the mode B at the upper band edge ($k = 0$) has a much larger $\text{Im}(\tilde{f})$ than that of mode A. This indicates that mode B has a larger radiaton damping, and its mode profile is plotted in Fig. 2(i). On the contrary, $\text{Im}(\tilde{f})$ of mode A at *k*=0 is almost zero ($< 10^{-7}$ THz), and its mode profile is shown in Fig. 2(h). It can be seen that the symmetry of mode A is antisymmetric about *x*=0, but mode B is symmetric about *x*=0. This illustrates that the mode profile of A does not match with that of the light in free space, and as such, is a BIC. Figures 2(a) and 2(d) show that all the nearby modes except A are radiative ones. Such symmetry always holds when $h_b$ is varying. Therefore, modes A, C, and F are all BICs at *k*=0, similar to the BICs that appeared in other optical systems[33], which can be utilized to manipulate thermal emission[48,49].

With the increase of $h_b$, both the contrast of $\text{Re}(n)$ and $\text{Im}(n)$ between these two multi-layer regions decrease monotonically. This provides two competitive mechanisms for changing the band gap. The decrease of $\text{Re}(n)$ contrast will reduce the gap, while the decrease of Im($n$) difference will enlarge the gap. Therfore, two bands will "meet" at a certain $h_b$, which is confirmed in Fig. 2(g). We find that at $h_b = 0.877$ μm, a pair of EPs appears at $k\Lambda/(2\pi) = \pm 0.0227$ and $\tilde{f} = (27.891 - 0.24857i)$ THz, which are marked as green dots. Note that at $k$=0, bound mode C and radiative mode D are orthogonal by symmetry, which can be seen from their mode profiles in Figs. 2(j) and 2(k). They cannot couple and the coupling coefficient must be zero at $k$=0. However, the coupling coefficient is non-zero at a finite $k$, and an EP can appear at a particular $k$-value if the coupling coefficient is equal to the difference in radiation damping. The EPs must come in plus and minus $k$ pair due to parity symmetry, and the gap closing points between such pairs of EPs are connected as the bulk Fermi arc in the non-Hermitian system, as shown by the black line in Fig. 2(g). This is because it links the two EPs with topological charges $\pm 1/2$ (see Section II of Supplementary Material)[42]. The bulk Fermi arc intersects with the BIC modes denoted by a solid red curve $\widehat{ACF}$ at C, as shown in the insert of Fig. 2(g). It is also worthy to mention that for smaller $h_b$, the less radiative band carrying the BIC at $k$=0 has lower Re($\tilde{f}$) in frequency, while for bigger $h_b$, the less radiative band has higher Re($\tilde{f}$) in frequency. The change in $h_b$ induces a topological band inversion since the symmetry and anti-symmetry modes swap their positions (see Section III of Supplementary Material).

## IV. Results

### A. Experimental Observation of BICs and EPs

The dispersion of meta-crystals is measured with ARTES. Experimental and the corresponding simulation results are compared in Fig. 3. It is not easy to directly calculate the far field thermal radiation, and hence the absorption spectra instead was calculated according to Kirchhoff's law[50]. Figures 3(a)–3(c) show the calculated optical absorption spectra using the simulation software, Lumerical FDTD, at different incident angles for three values of $h_b$ (see Supplementary Fig. S4 for details). To quantitatively compare with the measured result, the experimentally retrieved permittivity of gold and germanium in the finite-difference time-domain (FDTD) simulations were used. The measured thermal emission spectra are shown in Figs. 3(d)–3(f), which agree well with the FDTD results. Two bands can be clearly seen in the spectra, and their intensities vary dramatically with $h_b$. For $h_b = 0.816$ and $1.15$ μm in Figs. 3(a), 3(c), 3(d) and 3(f), the modes marked with red dots at $k = 0$ show very small radiations (or absorption). These modes correspond to the BICs displayed in Fig. 2. Around $h_b = 0.8784$ μm, a band inversion occurs, and the BIC transits from the lower to upper band, indicating a topological transition when the band gap closes. The spectra in Figs. 3(b) and 3(e) indicate that a pair of EPs appears at $k\Lambda/(2\pi) = \pm 0.0220$, directly at the transition point. For comparison, the real parts of eigenfrequencies $\text{Re}(\tilde{f})$ calculated by COMSOL are plotted as blue dashed lines in Figs. 3(a)–3(f). The good agreement between measurement and calculations corroborates the existence of a pair of EPs at a specific value of $h_b$.

Besides real parts of eigenfrequencies $\text{Re}(\tilde{f})$, the imaginary part $\text{Im}(\tilde{f})$ are required to determine both BICs and EPs. As the quality factors of the emission curves are related to $\text{Im}(\tilde{f})$ [shown in Figs. 2(d)–2(f)], $\text{Im}(\tilde{f})$ can be retrieved from the experimental spectra. Figures 3(g)–3(i) shows the emission spectra for three samples at wave vectors, $k\Lambda/(2\pi) = 0$ (purple dots) and $k\Lambda/(2\pi) = 0.06$ (red dots), which are correspondingly labelled by magenta and red dashed lines in Figs. 3(d)–3(f). For $k\Lambda/(2\pi) = 0.06$, there are two emission peaks originating from two photonic bands. However, for $k\Lambda/(2\pi) = 0$, only one emission peak appears due to the disappearance of the BIC emission peaks. To retrieve $\text{Im}(\tilde{f})$, the Lorentzian line shape functions are used to fit the emission spectra in Figs. 3(g)–3(i). The eigenfrequency were retrieved through fitting the measured thermal emission spectra using the function $I = I_0 + \left(\sum_j \frac{t_j}{f - (\text{Re}(\tilde{f}_j) + i\text{Im}(\tilde{f}_j))}\right)\left(\sum_j \frac{t_j}{f - (\text{Re}(\tilde{f}_j) + i\text{Im}(\tilde{f}_j))}\right)^*$. Here, $I$ is intensity, $I_0$ is the background intensity, $t_j$ is the complex fitting parameters, and $f$ is the experimental frequency. In Figs. 3(g)–3(i), the solid lines are fitted curves of Lorentzian line shape functions, which fit the experiment spectra quite well. The retrieved $\text{Im}(\tilde{f})$ are plotted as blue squares in Figs. 3(j)–3(l), which are compared with COMSOL calculations shown by blue dashed lines. The extracted $\text{Im}(\tilde{f})$ from measurements agree with COMSOL calculations. Especially, the extracted $\text{Im}(\tilde{f})$ in Fig. 3(k) confirms the EP at $h_b = 0.8755$ μm depicted in Figs. 3(b) and 3(e). If the $\text{Im}(\tilde{f})$ of three samples in Figs.

3(j)–3(l) are compared, the absolute values of $\text{Im}(\tilde{f})$ decrease with the increase of $h_b$. This is because, for larger $h_b$, guided modes can be better confined inside the meta-crystal and the radiation loss is reduced, leading to smaller absolute values of $\text{Im}(\tilde{f})$.

### B. BIC Line and Fermi Arc in Synthetic Dimensions

In order to compare the above experiments with theoretical calculations in Fig. 2(g), Fig. 4(a) shows measured real parts of eigenfrequencies $\text{Re}(\tilde{f})$ in 2D synthetic parameter space $(k, h_b)$. The color of each dot represents the inverse of experimentally extracted $|\text{Im}(\tilde{f})|$. The results of 8 samples with different values of $h_b$ are shown in Fig. 4(a), and their experimental and simulation details are given in supplementary Fig. S5. Most dots are in blue, but there exists a string of red dots at $k\Lambda/(2\pi) = 0$, corresponding to BICs mandated by symmetry. However, the color contrast between BICs and other radiative modes in Fig. 4(a) is smaller than that in Fig. 2(g), because ohmic losses inside the gold grating [which is not included in Fig. 2(g)] contribute to $\text{Im}(\tilde{f})$ and reduce quality factors of BIC in the experiment. Nevertheless, in Fig. 4(a), the $|\text{Im}(\tilde{f})|$ of BICs are still local minima of emission. Varying $h_b$ induces a band inversion of two bands at $h_b \cong 0.878$ μm. At this transition point, a bulk Fermi arc connects two EPs, shown as the solid black line in the 2D synthetic parameter space in Fig. 4(a). The BIC line intersects with the bulk Fermi arc at $k\Lambda/(2\pi) = 0$ and $h_b \cong 0.878$ μm in the synthetic parameter space.

### C. Exceptional Line in Synthetic Dimensions

In the above discussions, a pair of EPs are shown at one particular value of $h_b$. Actually, more EPs can be found in a higher dimensional space. In Fig. 1(b), besides $h_b$, the thickness $h_a$ can also be used as another synthetic parameter. In the new design, $h_a$ and $h_b$ are chosen as independent tunable parameters and other parameters ($d_b$, $\Lambda$, and $h_d$) are fixed. A 3D synthetic parameter space is constructed using $(k, h_a, h_a + h_b)$. Through scanning these parameters in COMSOL simulations, all the EPs obtained in this 3D synthetic space form ELs, which are shown as black dashed lines in Fig. 4(b). As the EPs always appear in pairs at two wave vectors $k$ and $(-k)$ in reciprocal systems, there are two ELs which are symmetric with respect to the $k=0$ plane. The two dashed lines are degenerated at $k\Lambda/(2\pi) = 0$, indicating the pair of EPs originating from the degeneracy at the $\Gamma$ point. In the experiment, four samples are fabricated, and their EPs can be determined by ARTES [represented as dots in Fig. 4(b)]. The results of sample $\alpha$ are already provided in Fig. 3, and the results of other three samples ($\beta, \gamma, \delta$) are provided in Supplementary Fig. S6. All the measured EPs are well located at the calculated dashed line, implying a good agreement between calculations and measurements.

### V. Conclusions

We designed and experimentally realized metallic meta-crystals that can work as a flexible system to investigate various non-Hermitian and topological physics. The meta-crystal exhibits symmetry protected BICs. Through changing the thickness of the cover layer above the meta-crystal, the radiative losses of meta-crystals can be controlled to induce band inversions. At the crossover point, EPs and the associated non-Hermitian Fermi arc can be observed linking them. Various features in synthetic parameter spaces, such as the trajectory of BICs and bulk Fermi arcs, can be mapped using the ARTES platform with a high angular resolution. The interesting physics studied here can be utilized to manipulate thermal emission, including emission spectra and radiation patterns. For example, the BIC can suppress the thermal emission in one particular direction, which can be useful in thermophotovoltaic devices and radiative cooling. In addition, as optical absorption is the reverse process of thermal emission (from the Kirchhoff law point of view), this scheme can also be used to control optical absorption of light coming from the far field.

More importantly, this work demonstrates the usefulness of a home-made ARTES platform which offers a simple but powerful experimental technique to characterize the dispersions of non-Hermitian systems. It has an excellent angular resolution and operates in a broad frequency range, from near to far-infrared. It will be a useful tool to explore the topological band structures of non-Hermitian photonic crystal slabs consisting of various microstructures types.

**Acknowledgements**

This work was financially supported by the National Natural Science Foundation of China (Nos. 11690033, 61425018 and 11621091), National Key Research and Development Program of China (No. 2017YFA0205700) and National Key R&D Program of China (2017YFA0303702). Work done in Hong Kong was supported by Research Grants Council (RGC) Hong Kong through grants N_HKUST608/17 and AoE/P-02/12. K.D. acknowledges funding from the Gordon and Betty Moore Foundation.

**Author contributions**

F.Z., K.D. and Y.Z. contributed equally to this work. F.Z. Y.Z. H.L. did the experiments, including fabrication and measurement. C.T.C. and S.N.Z supervised the project. All authors contributed to the writing of manuscript.

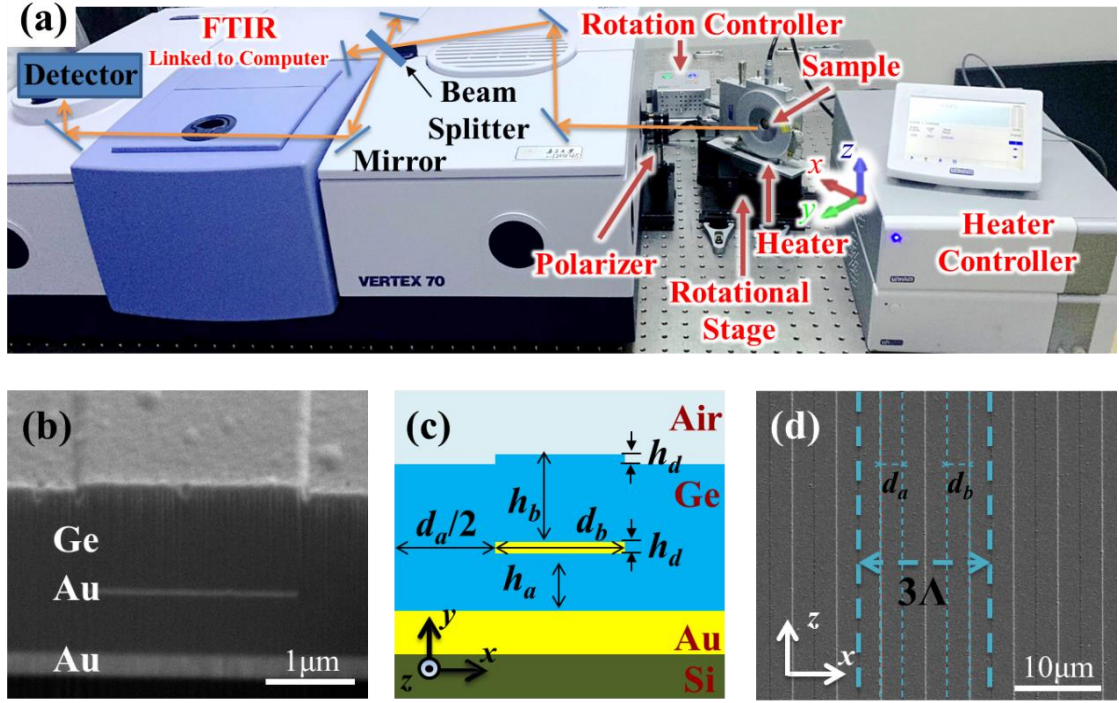

FIG. 1. ARTES setup and structure of the meta-crystal. (a) Pictures and schematics of the ARTES setup. The sample is heated inside the heater with a window to output the signals. The heater is placed on a precisely controlled electric rotational stage. (b) SEM side-view of one sample. The dark region is germanium (Ge), and the gray strip in the middle is the gold grating. (c) Schematics of the unit cell with structural parameters $h_a$, $h_b$, $h_d$, and $\Lambda = d_a + d_b$. (d) SEM top-view of a sample with structure parameters $d_a$, $d_b$ and $\Lambda$, marked by the cyan dashed lines.

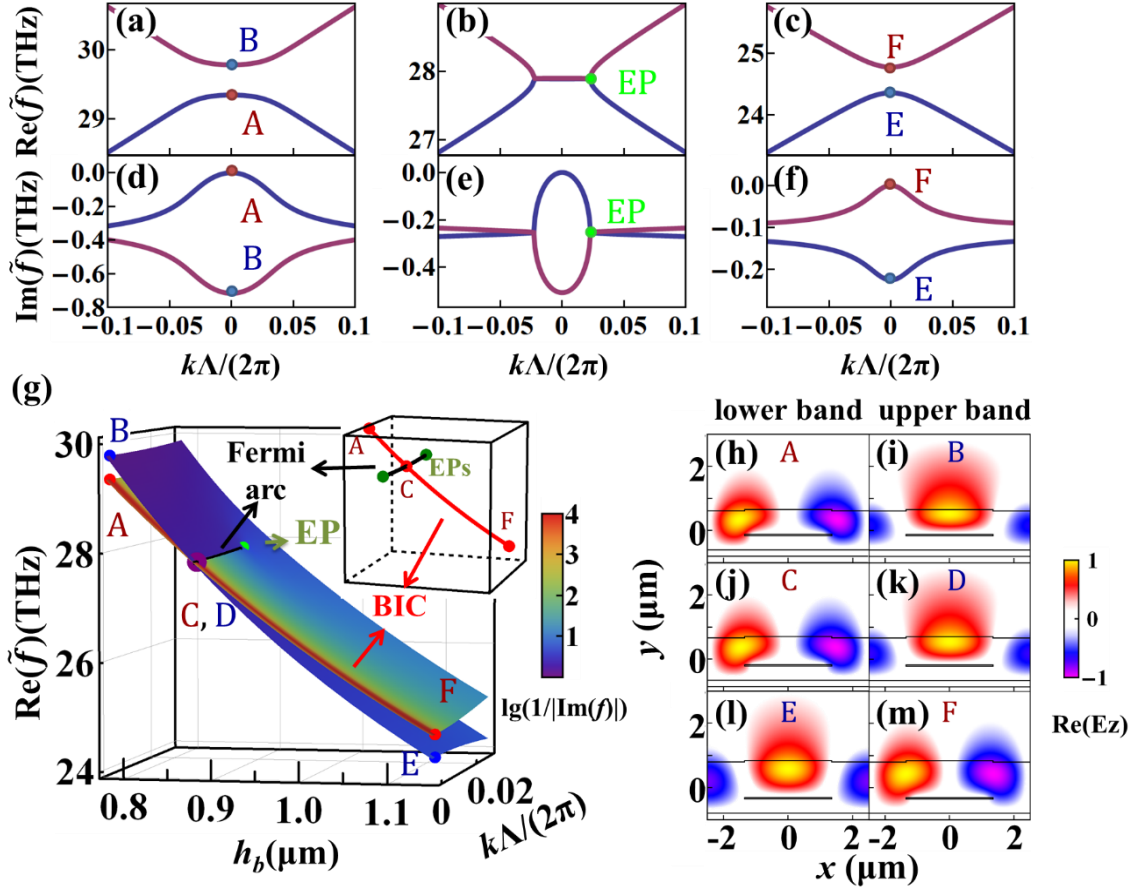

FIG. 2. Calculated band structures and eigenmode profiles. Calculated dispersion of (a)–(c) real and (d)–(f) imaginary frequencies with different $h_b$ parameters: (a), (d) $h_b = 0.78$ μm; (b), (e) $h_b = 0.877$ μm; (c), (f) $h_b = 1.15$ μm. Au is described by a lossless Drude model $\varepsilon_{Au} = 1 - \omega_p^2/\omega^2$ ($\omega_p = 9.0$ eV) and $\varepsilon_{Ge} = 16.04$. Structural parameters are $\Lambda = 5$ μm, $d_a = 2.7$ μm, $d_b = 2.3$ μm, $h_a = 0.44$ μm, and $h_d = 45$ nm. The three cases in (a)–(f) correspond to the three cutting slices in (g). (g) Dispersion of real part of eigenfrequencies $Re(\tilde{f})$ of TE-modes in the synthetic parameter space $(k, h_b)$, calculated with COMSOL. The color of the surface represents the inverse of imaginary parts of eigenfrequencies $Im(\tilde{f})$ in the log-10 scale. The black line denotes the modes with identical real part of eigenfrequencies (also referred as bulk Fermi arc in the non-Hermitian system) linking up two EPs (green dots). The inset shows the crossing between the BIC (red line) and the bulk Fermi arc (black line). (h)–(m) The Ez field profile of eigenmodes with their parameters $(Re(\tilde{f}), k\Lambda/(2\pi), h_b)$ given as A (29.349 THz, 0, 0.78 μm), B (29.785 THz, 0, 0.78 μm), C (27.852 THz, 0, 0.8797 μm), D (27.852 THz, 0,

0.8797 µm), E (24.357 THz, 0, 1.15 µm), and F (24.774 THz, 0, 1.15 µm).

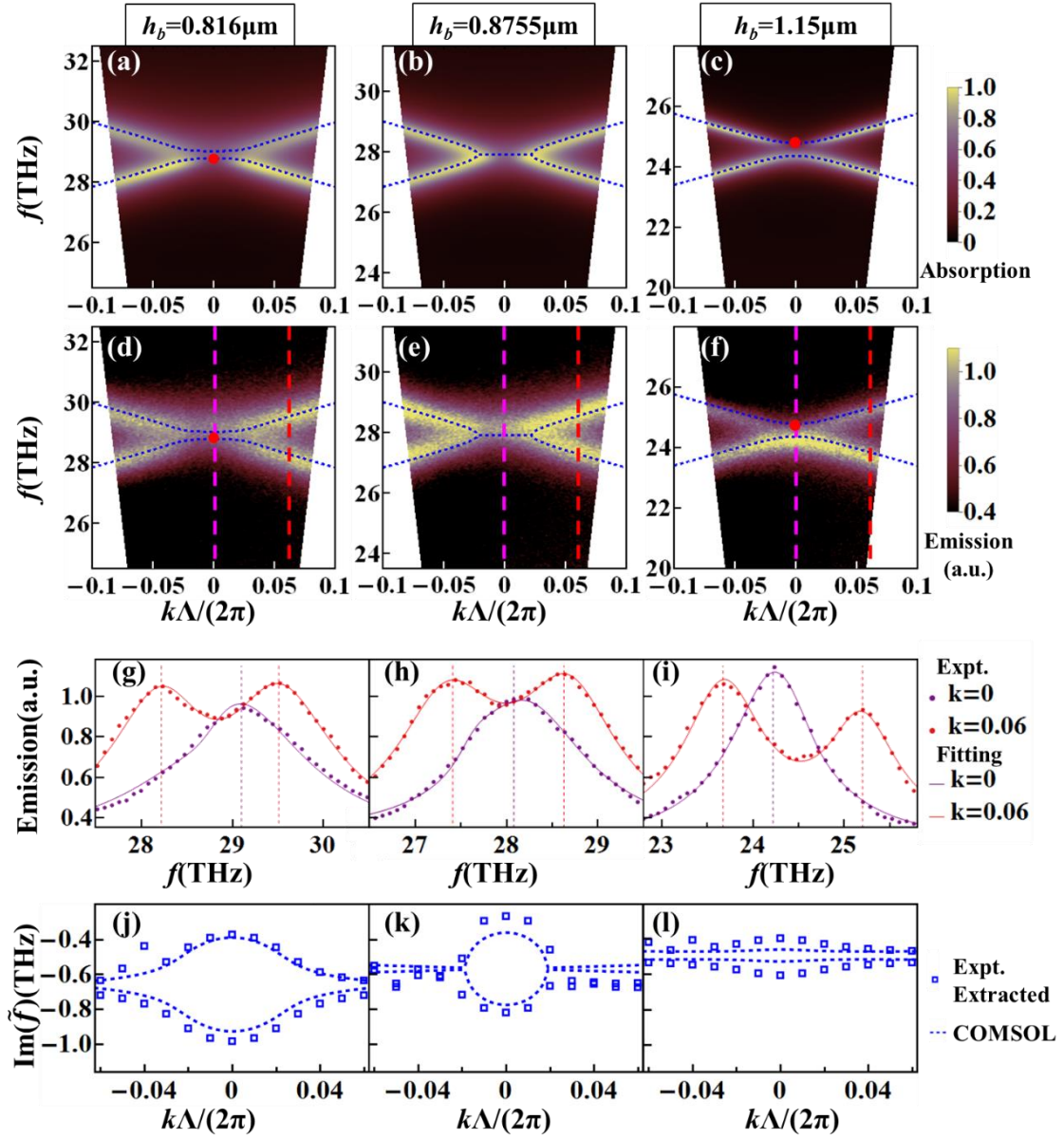

FIG. 3. Experimental observation and numerical simulations of EPs and BICs. (a)–(c) Optical absorption simulated by FDTD with realistic material parameters (Au-Palik and Ge-Palik from FDTD material database) and (d)–(f) measured thermal emission: (a), (d) $h_b = 0.816$ μm; (b), (e) $h_b = 0.8755$ μm; (c), (f) $h_b = 1.15$ μm. The blue dashed lines in (a)–(f) are calculated eigenfrequencies by COMSOL with the Drude model ($\varepsilon_{Au} = 1 - \omega_p^2/(\omega^2 + i\omega\omega_c)$, $\omega_p = 9.0$ eV and $\omega_c = 0.027$ eV). Here, the structure parameters are $\Lambda = 5$ μm, $d_a = 2.7$ μm, $d_b = 2.3$ μm, $h_a = 0.44$ μm, and $h_d = 45$ nm. The measurement resolution in (d)–(f) is $\delta k \sim \delta\theta/\lambda \approx 0.02$ degree/μm. (g)–(i) The solid dots are thermal emission spectra with wave vectors marked by the vertical dashed lines in (d–

f) at $k\Lambda/(2\pi) = 0$ (purple dots) and $k\Lambda/(2\pi) = 0.06$ (red dots), and the corresponding solid lines are the fitted curves to the dots. In (g)–(i), all the purple lines have one resonance peak, and the red curves have two resonance peaks. The resonance frequency of the peak in the purple line is redshifted with increasing $h_b$. In (j)–(l), blue square dots give the imaginary part of eigenfrequencies extracted from experimental data in (d)–(f). The blue dashed lines are calculated by COMSOL with the Drude model [$\omega_p = 9.0$ eV, $\omega_c = 0.162$ eV in (j) and (k), and $\omega_c = 0.324$ eV in (l)].

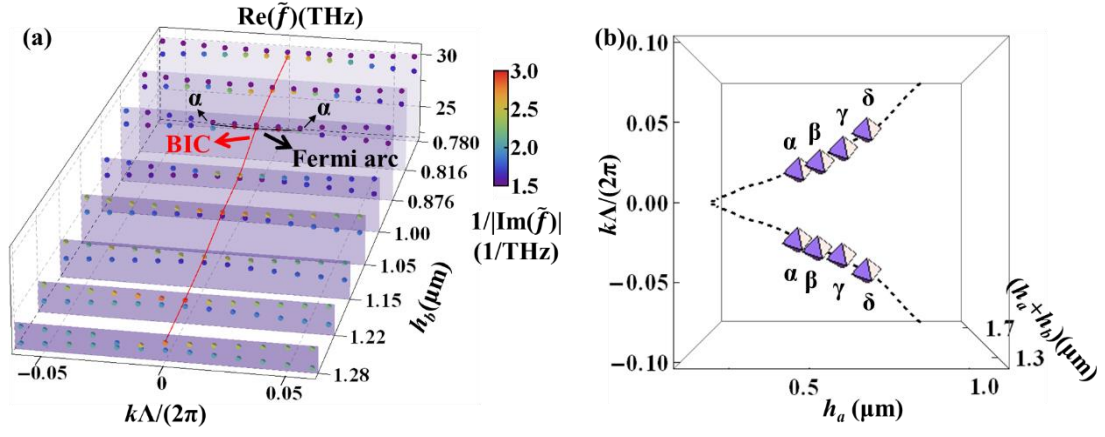

FIG. 4. BIC, Fermi arc and exceptional lines in synthetic space. (a) Real parts of eigenfrequencies in 2D parameter space ($k$, $h_b$), which are read from experimental data in Fig. 3. The data are plotted as filled dots with their colors denoting different imaginary parts of eigenfrequencies. The BIC line and bulk Fermi arc calculated by COMSOL are shown as the red and black curves, respectively. (b) Exceptional lines in 3D parameter space ($k$, $h_a$, $h_a + h_b$). Here, the parameters are set as $d_b = 2.3$ μm, $\Lambda = 5$ μm, $h_d = 45/440 \cdot h_a$, while $h_a$ and $h_b$ are variables. The measured EP pairs are represented as dots marked by α, β, γ, δ. In the experiment, four samples are fabricated with their parameters ($h_a$, $h_a + h_b$) given as α (0.44, 1.3155 μm), β (0.51, 1.3732 μm), γ (0.6, 1.4575 μm), δ (0.7, 1.5242 μm).